# Negative spin Hall magnetoresistance in antiferromagnetic $Cr_2O_3$/Ta bilayer at low temperature region

Yang Ji[1], J. Miao,[1,a)] Y. M. Zhu[1], K. K. Meng,[1] X. G. Xu,[1] J. K. Chen[1], Y. Wu,[1] and Y. Jiang[1,b)]

*School of Materials Science and Engineering, University of Science and Technology Beijing, Beijing, 100083, China*

**ABSTRACT**

We investigate the observation of negative spin Hall magnetoresistance (SMR) in antiferromagnetic $Cr_2O_3$/Ta bilayers at low temperature. The sign of the SMR signals is changed from positive to negative monotonously from 300 K to 50 K. The change of the signs for SMR is related with the competitions between the surface ferromagnetism and bulky antiferromagnetic of $Cr_2O_3$. The surface magnetizations of α-$Cr_2O_3$ (0001) is considered to be dominated at higher temperature, while the bulky antiferromagnetics gets to be robust with decreasing of temperature. The slopes of the abnormal Hall curves coincide with the signs of SMR, confirming variational interface magnetism of $Cr_2O_3$ at different temperature. From the observed SMR ratio under 3 T, the spin mixing conductance at $Cr_2O_3$/Ta interface is estimated to be $1.12 \times 10^{14}$ $\Omega^{-1} \cdot m^{-2}$, which is comparable to that of YIG/Pt structures and our early results of $Cr_2O_3$/W. (Appl. Phys. Lett. 110, 262401 (2017))

---

a) Electronic mail: j.miao@ustb.edu.cn
b) Electronic mail: yjiang@ustb.edu.cn

The resistance of ferromagnetic materials (FMs) change depending on the magnitude of the external magnetic field, which is called magnetoresistance (MR). So far giant magnetoresistance (GMR),[1,2] anisotropic magnetoresistance (AMR)[3] and tunnel magnetoresistance (TMR)[4-7] etc. have been discovered in the recent decades, but those MRs need a condition that the current must travel through the FMs, so these phenomenons only exist in conductive FMs. Moreover, spin Hall magnetoresistance (SMR) is a special MR that there is no charge current in the FM layer, which depends on the relative orientation between the magnetic moments of FM layer and the spin direction injected from normal metal (NM).[8] H. Nakayama *et al*. first observed 0.01% MR ratio in YIG/Pt structure, and defined as SMR.[9] Until now, SMR has been reported for NM/ ferromagnetic insulators such as YIG,[9-12] $Fe_3O_4$,[13] $NiFe_2O_4$,[14] and $CoFe_2O_4$[15], etc..

It is known that the spin current plays an important role in the dynamics of SMR injecting from NM layer into magnetic layer. Due to the spin Hall effect (SHE), when the charge current flow through the NM layer, a spin current is generated in the vertical direction and pass the interface to arrive in FM layer. Meanwhile, the orientation relationship between the direction of the magnetic moment (**m**) in the magnetic layer and the polarization direction (**s**) of the spin current injected will determine the magnitude of the charge current by the inverse spin Hall effect (ISHE), which shows different resistance of NM layer, and lead to SMR.[8,9]

In recent years, antiferromagnetic spintronics attract much research attentions.[16,17] Han *et al.* detected SMR signal in the antiferromagnetic insulator $SrMnO_3$.[18] Similarly, SMR has also been reported in antiferromagnetic metallic FeMn/Pt.[19] A. Manchon investigated physical origin of SMR in the antiferromagnets and predicted that SMR can be observable in other antiferromagnetic insulators such as NiO, CoO, $Cr_2O_3$ etc.[20] In previous work, we observed a nearly 0.1% SMR ratio in $Cr_2O_3$/W with 9 T.[21] Soon after that, a negative SMR has also been observed in NiO bulk crystal and films.[22-24]

As known, $Cr_2O_3$ is insulating magnetoelectric antiferromagnet, in which the magnetism can be controlled by electric field. [25] Due to its Néel temperature is 308 K, namely above the room temperature, which makes $Cr_2O_3$ become potential candidate material in antiferromagnetic spintronics. Recently T. Kosub put forth an concept of antiferromagnetic magnetoelectric random access memory, in which the prototypical memory cell consists of an active layer of $Cr_2O_3$.[27] Relying on its advantages, more spintronic phenomenon will occur in $Cr_2O_3$ magnetoelectric. However, the physical mechanism in $Cr_2O_3$/heavy-metal structure has not been investigated in details.

In this work, we investigated the temperature dependence of SMR in a $Cr_2O_3$ (25)/Ta (5) bilayer, and confirmed that negative SMR is originated from the magnetizations of $Cr_2O_3$. The (0001)-oriented $Cr_2O_3$ films were grown on the (0001) oriented rutile $Al_2O_3$ substrates via pulse laser deposition (PLD) with a base pressure of $5 \times 10^{-8}$ mbar. Prior to the $Cr_2O_3$ growth, the substrate temperature was increased to 450 °C with a rate of 10 °C/min, and the thin film deposition was performed with an oxygen partial pressure of 0.05 mbar. Then the $Cr_2O_3$ film was deposited with a laser power of 1.8 J/cm$^2$ and a frequency of 3.0 Hz and the target to substrate distance was maintained at 5 cm during the deposition. The cooling process was carried under $10^6$ oxygen partial pressure with a rate of 10 °C/min to room temperature. After deposition, a 5 nm film Ta was grown *in situ* on the $Cr_2O_3$ by magnetron sputtering, where the base pressure of the chamber was less than $8 \times 10^{-8}$ mbar.

Finally, a hall bar for electrical measurements was fabricated by electron beam lithography and Ar ion etching.

The optical image and experimental hall geometry of Cr$_2$O$_3$/Ta bilayers are shown in Fig. 1(a). The size of Hall-bar is 400 μm×40 μm and a constant channel current **I** of 20 μA along the **x** direction, in which the Hall-bar structure was patterned onto the substrate. The phase structure of the Cr$_2$O$_3$ films was determined by X-ray diffraction using M21XVHF22 X-ray diffractometer with Cu/Ka. Fig. 1(b) shows the XRD ω-2θ scan for the Al$_2$O$_3$ (0001)/Cr$_2$O$_3$ (25 nm) sample. Obviously Cr$_2$O$_3$ (0006) and Cr$_2$O$_3$ (00$\underline{1}$2) peaks can be observed next to the substrate peak at 39.8° and 86.1°, respectively, which demonstrates uniaxial orientation growth of α-Cr$_2$O$_3$.[25] In addition, the surface morphologies was checked by using a scanning probe microscopy in Fig. 1(c) and the root-mean-squar roughness is 0.217 nm for 25 nm Cr$_2$O$_3$ film. The surface of Cr$_2$O$_3$ layer is relatively smooth without any cracks or pinholes, which is beneficial to growing the upper heavy metal Ta layer.

According to the SMR theory,[8,9] a charge current flowing along the x direction in Ta layer generates a y-polarized spin current, flowing along the z direction and injecting into Cr$_2$O$_3$ layer. Depending on the orientation of moments of Cr$_2$O$_3$ with respect to the spin direction of spin current, the reflected spin current varies in the magnitude, which yields an additional charge current via ISHE superimposed on the original charge current. as known, the SMR is described as [8]

$$\rho_{xx} = \rho_0 - \Delta\rho m_y^2, \qquad (1)$$

$$\frac{\Delta\rho_1}{\rho} = \theta_{SH}^2 \frac{\lambda}{d_N} \frac{2\lambda G_r \tanh^2\frac{d_N}{2\lambda}}{\sigma + 2\lambda G_r \coth\frac{d_N}{\lambda}}, \qquad (2)$$

where $\rho_0$ is a constant resistivity offset, $\mathbf{m_y}$ is the y component of the magnetization unit vector, and $\Delta\rho_1/\rho$ depends on spin diffusion length $\lambda$, Ta layer thickness $d_N$, Ta electrical conductivity $\sigma$, spin hall angle $\theta_{SH}$, and the real part of the spin mixing conductance $G_r$, as shown in Eq. (2).

Clearly, the in-plane field sweep cannot distinguish AMR from SMR since both signals depend on the orientation of moments in the **xy** plane. In order to extract the SMR contribution from the overall MR, both the field-dependent magnetoresistance (FDMR) measurements and angle-dependent magnetoresistance (ADMR) measurements were performed on the Cr$_2$O$_3$/Ta bilayer. As illustrated in Fig. 2(a), the longitudinal resistance of the sample was measured under rotating a constant field H = 3 T with α, β, γ, respectively. The different ADMR would reveals different physical mechanisms in two cases. i) if magnetic moments are rotated in the xz plane followed by angle γ, SMR should remain constant, since $\mathbf{m_y}$ and the reflected spin current are unchanged, namely any resistance change can be attributed to AMR. ii) if moments are rotated in the **yz** plane followed by angle β, AMR should remain constant, since the charge current is always perpendicular to the moments, namely any resistance change can be attributed to SMR. However, if moments are rotated in the **xy** plane followed by angle α, both SMR and AMR change simultaneously and therefore the two MR effects are entangled.

Fig. 2(b) shows FDMR measurement at 300 K. Clearly, we can observe the MR in x-axis is the highest, and the MR in y-axis is the smallest. It should be noted that the magnetization of the bilayers is difficult to saturate, even the field is increased to 3 T. Fig. 2(c) shows the ADMR measurement with 3 T at 300 K, in which MR dependence on β-angle represents SMR and MR dependence on γ-angle represents AMR, respectively. A conventional positive SMR can be observed and the SMR ratio is $(R_z - R_y)/R_z = 2.4\times 10^{-4}$. Similarly, FDMR and ADMR measurements at 50 K are exhibited in Fig. 2(d) and Fig. 2(e). Interestingly, Comparing with MRs at 300 K, both FDMRs and ADMRs reverse their signs at 50 K, in which a negative SMR can be observed clearly. The ratio of SMR in $Cr_2O_3$/Ta bilayer is about $0.5\times 10^{-4}$, which is close to that of NiO/Pt structure.[24]

To confirm those observations, Fig. 3(a) shows the longitudinal resistance $R_{xx}$ dependence on β-angle for $Cr_2O_3$/Ta bilayer with 3 T at different temperature. The corresponding temperature dependence of SMR were described in Fig. 3(b). Unambiguously, as temperature decreasing, SMR varies from positive to negative monotonically. When the sample was cooled down to 250 K, the positive SMR would decrease to zero sharply. As the temperature decreased further, the negative SMR gradually emerged. Nevertheless, the slope of the SMR curve starts to slow down below 250 K, and the ratio of SMR reaches a stable value until 50 K.

To investigate its physical origination, the magnetization structure of $Cr_2O_3$ need to be taken into account. As known, a boundary magnetization exists at the surface of the magnetoelectric antiferromagnet (0001) $Cr_2O_3$.[25,26] Therefore, as shown in Fig. 4(a), the magnetisms of $Cr_2O_3$ can be discriminated into two parts: surface ferromagnetism and bulk antiferromagnetism. Both of two parts will make contributions to SMR of $Cr_2O_3$/Ta. When the temperature is higher than 250 K, the antiferromagnetic order in $Cr_2O_3$ is weaker and its corresponding contribution to SMR is less than that from surface ferromagnetism. Thus, the sign of SMR shows positive, like FM/NM structure such as YIG/Pt.[9-12] Oppositely, when the temperature is below 250 K, the antiferromagnetic order gets strong and surface ferromagnetism gets weaker. Thus, the contributions from antiferromagnetic order is more than that contributed from surface ferromagnetism to SMR. Therefore, the sign of SMR in Cr$_2$O$_3$/Ta shows negative. A similar phenomenon has been reported in NiO/Pt structure.[22-24]

It should be noted that only the β-phase of Ta has a large spin Hall angle and strong spin-orbit coupling than other phase of Ta, so $\rho_{xx}$ as a function of temperature for a 5-nm-thick Ta/Cr$_2$O$_3$ film sample was measured. As shown in Fig. 4(b), the high resistivity of β-phase Ta is agreed with that reported increases with decreasing temperature.[28,29] Furthermore, from the equation (2), a higher SMR ratio with help of β-Ta can be obtained. At the same time, the spin mixing conductance $G_r$ can be estimated via the SMR ratio $\Delta\rho_1/\rho$ = 0.04 % with 3 T at 300 K. For Ta,[10] $\theta_{SH}$ = 0.02, λ = 1.8 nm, $d_N$ = 5 nm, and consequently the spin mixing conductance $G_r = 1.12 \times 10^{14}$ $\Omega^{-1} \cdot m^{-2}$. Due to the MRs are unsaturated till 3 T, the values of G$_r$ should be larger with increasing the magnetic field.

To confirm that explanations, the Hall resistance measurements were carried out in Cr$_2$O$_3$/Ta bilayer at different temperature, which are shown in Figs. 5(a) and 5(b). It should be noted that the transition temperature of Cr$_2$O$_3$ film is 250 K. i) when the temperature is higher than 250 K, the slopes of Hall curves are all positive and the abnormal Hall effect (AHE) can be observed. In addition, AHE signals of Cr$_2$O$_3$/Ta bilayer get stronger with increasing temperature, which is attributed to the surface ferromagnetism of Cr$_2$O$_3$. ii) when the temperature is lower than the transition temperature, the slopes of Hall curves are negative and no AHE signals can be observed. In other words, surface ferromagnetism may vanish with temperature decreasing, namely the bulk antiferromagnetic may exceed the surface ferromagnetism in the Cr$_2$O$_3$ thin films.[25]

In conclusion, we observed a negative SMR in $Cr_2O_3$/Ta bilayers below 250 K, which is attributed to the competitions between the surface ferromagnetism and bulky antiferromagnetics of $Cr_2O_3$. Above the transition temperature, the surface ferromagnetism of $Cr_2O_3$ is dominated, leading to a normal positive SMR. While the temperature is below the transition temperature, antiferromagnetic Néel order of $Cr_2O_3$ is robust and surface ferromagnetism makes less contributions to SMR, resulting in a negative SMR. The observations of negative SMR in antiferromagnet, like $Cr_2O_3$, NiO, may pave a way for applications in antiferromagnetic spintronics devices.

# ACKNOWLEDGEMENTS

This work was partially supported by the National Basic Research Program of China (Grant No. 2015CB921502), the National Science Foundation of China (Grant Nos. 11574027, 51731003, 51671019, 51471029) and Beijing Municipal Science and Technology Program (Z161100002116013) and Beijing Municipal Innovation Environment and Platform Construction Project (Z161100005016095). J.C., K.M., and Y.W. acknowledge the National Science Foundation of China (Nos. 61674013, 51602022, 61404125, 51501007).

**FIGURE LEGENDS**

Fig. 1 (a) Optical image of the $Cr_2O_3$/Ta bilayer surface and a schematic illustration of the electric resistance measured by the four-probe method, in which the length is 400 μm, and the width is 40 μm. (b) XRD 2θ scan from the 25 nm $Cr_2O_3$ film grown on $Al_2O_3$ (0001) substrate. (c) of 25 nm $Cr_2O_3$ film and the $R_q$ is 0.217 nm.

Fig. 2 (a) Notations of different rotations of the angular α, β, and γ. (b) and (d) show external magnetic field dependence of resistance curve for $Cr_2O_3$ (25)/Ta (5) at 300 K (b) and at 50 K (d). (c) and (e) show α, β, and γ dependence of resistance curve for $Cr_2O_3$ (25)/Ta (5) with 3 T magnetic field at 300 K (c) and at 50 K (e).

Fig. 3 (a) β dependence of the resistance in $Cr_2O_3$ (25)/Ta (5) with 3 T magnetic field at different temperature. (b) Temperature dependence of SMR signals in $Cr_2O_3$ (25)/Ta (5) under 3 T.

Fig. 4 (a) Schematic of $Cr_2O_3$ spin structure at T < $T_{Néel}$ in an AFM single domain state (lower two layers of arrows represent AFM structure of $Cr^{3+}$ spins in the bulk), which is accompanied by a positive boundary magnetization (top layer representing the surface). (b) Resistivity as a function of temperature of a representative 5 nm thick Ta Hall bar/$Cr_2O_3$ film by the four-probe method. Inset: the schematic illustration of measurement.

Fig. 5 AHE measurements of $Cr_2O_3$ (25)/Ta (5) at 50 -265 K (a) and 270 -300 K (b)

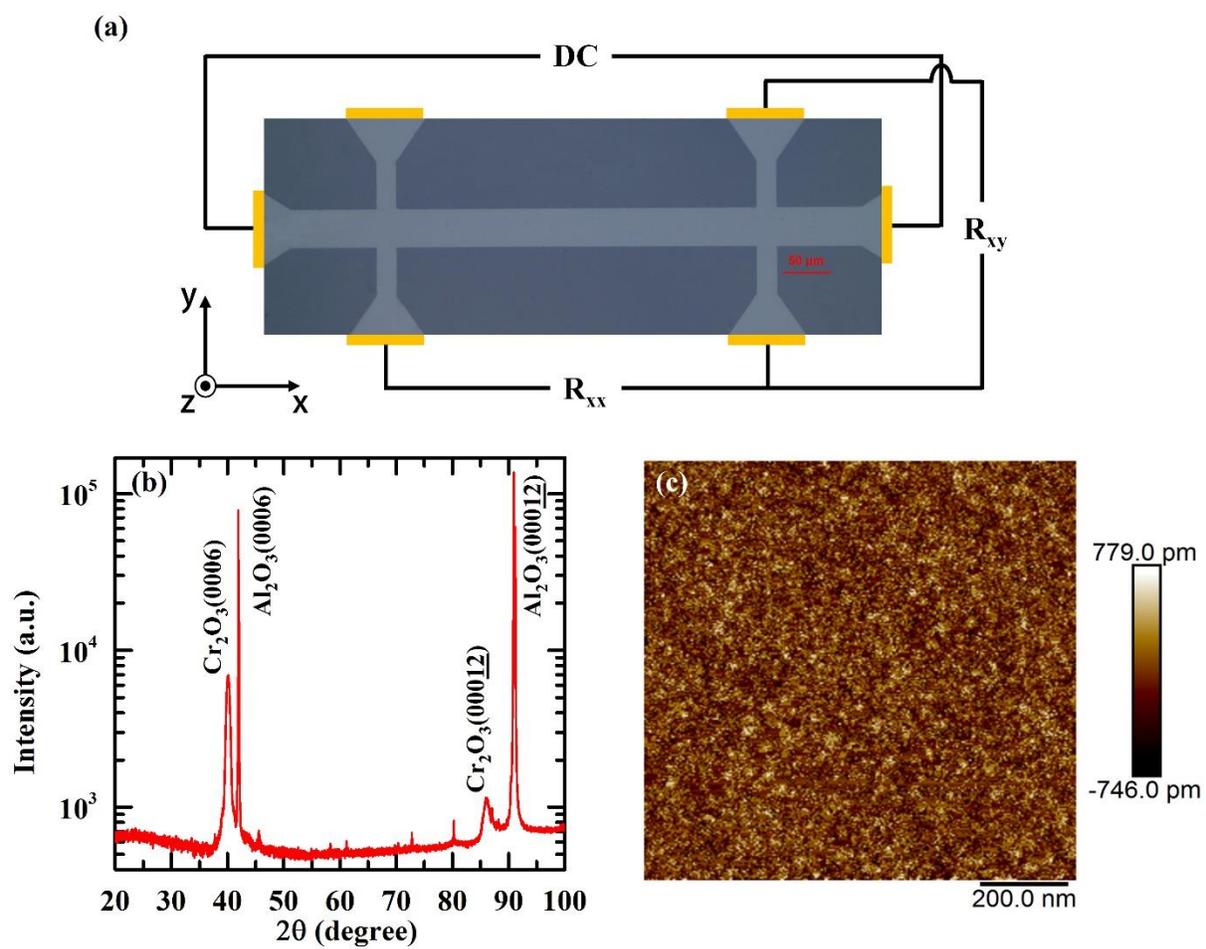

*Fig. 1*

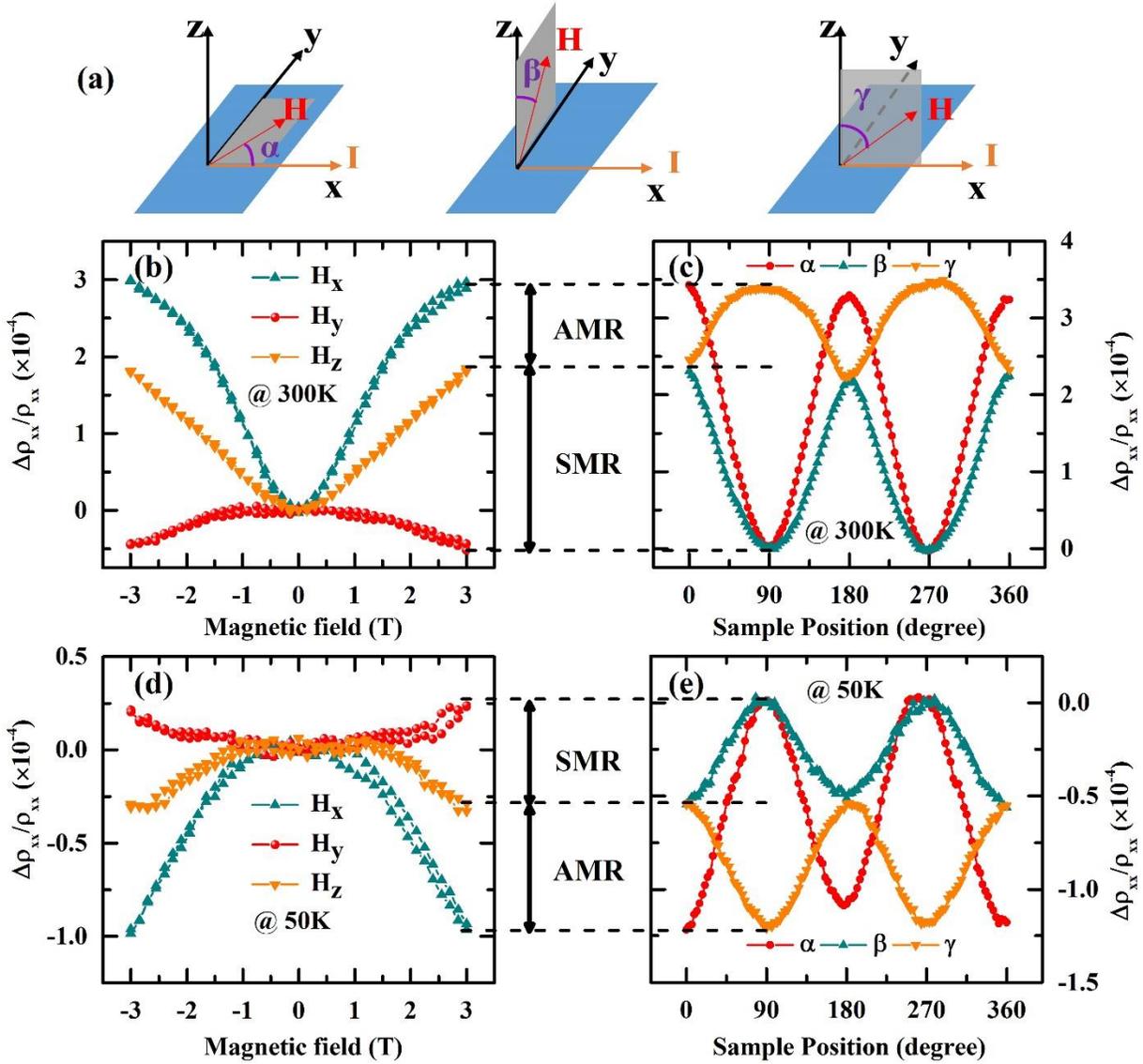

*Fig. 2*

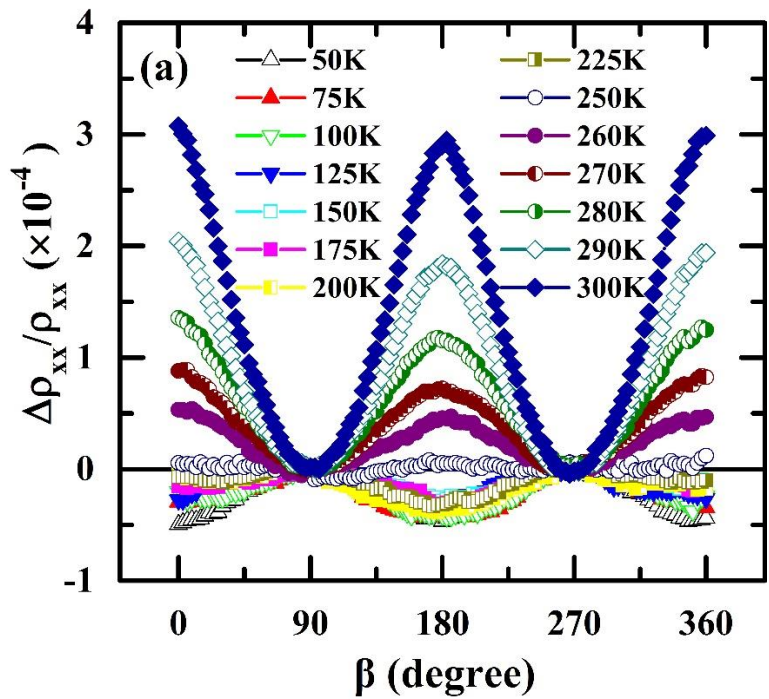
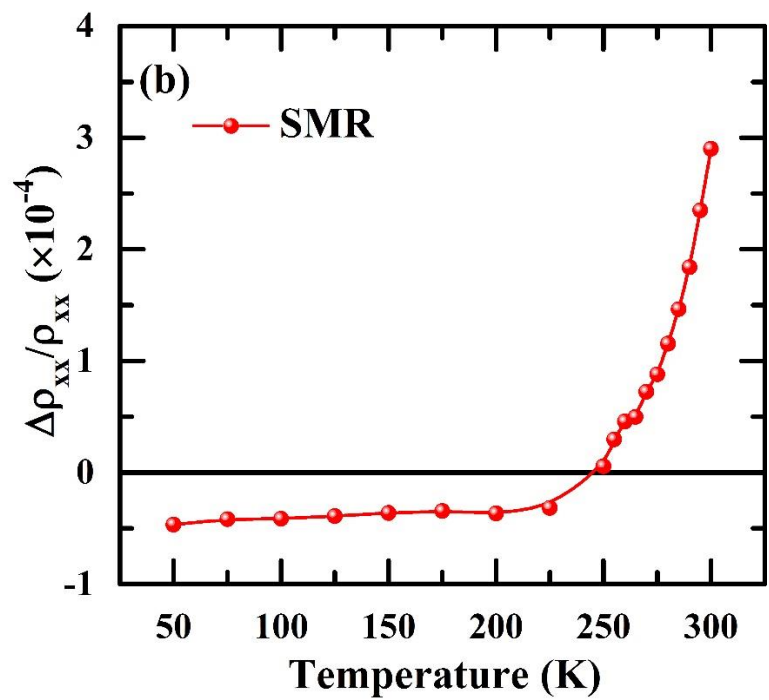

*Fig. 3*

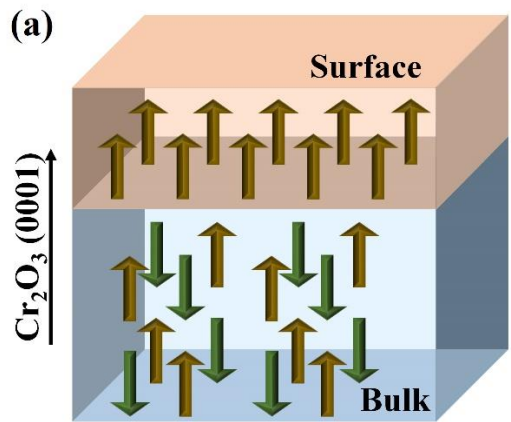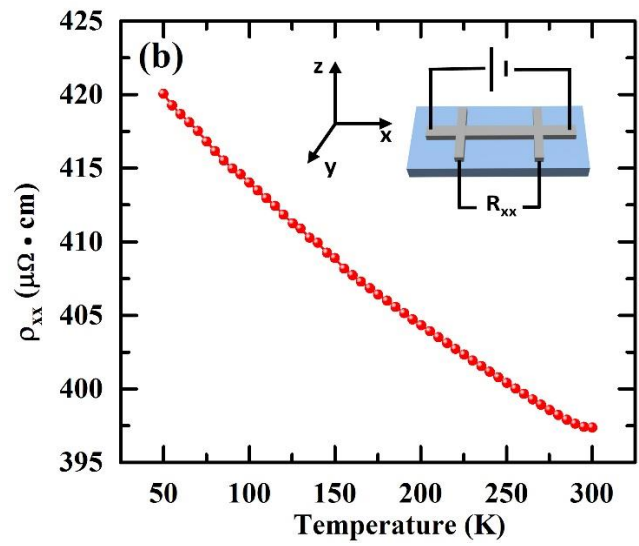

*Fig. 4*

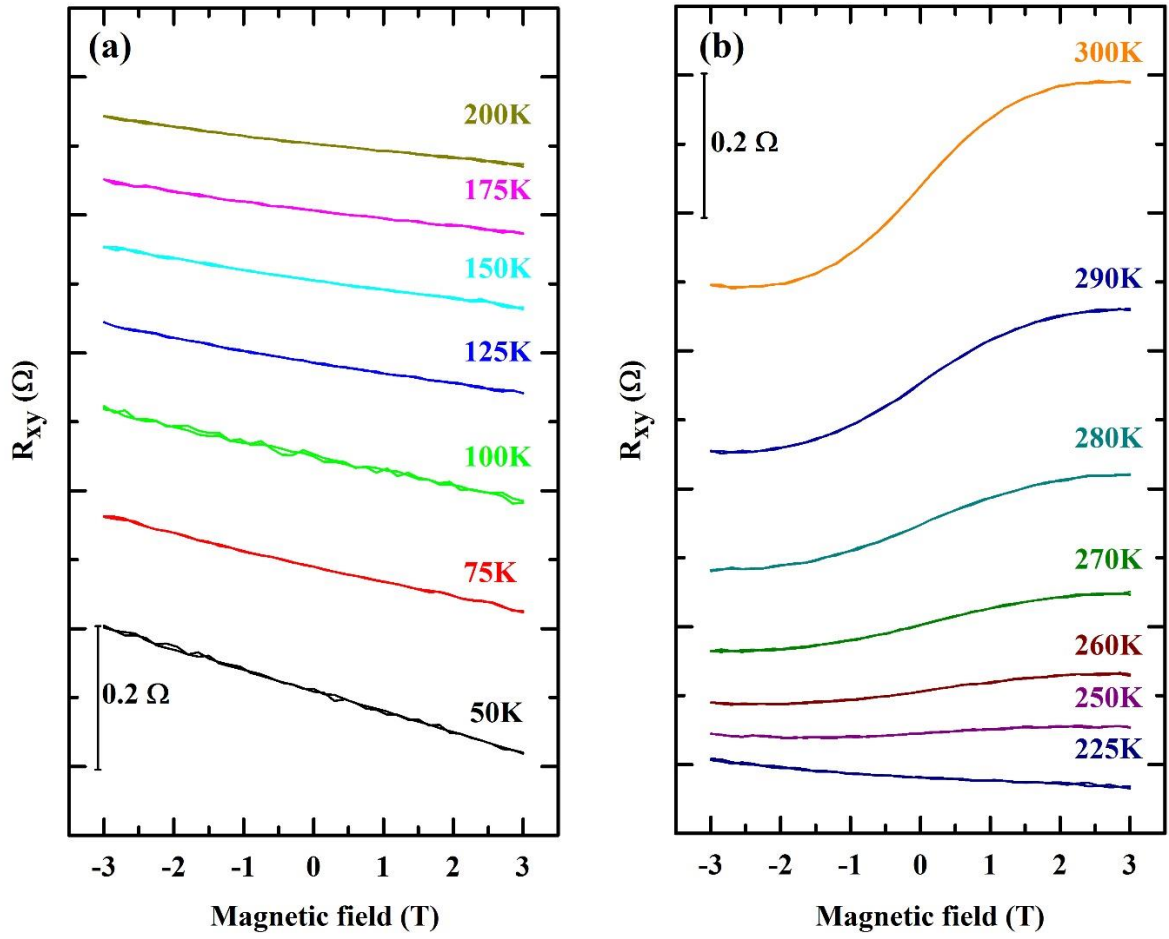

*Fig. 5*